\documentclass{aa}

\begin{document}

\thesaurus{06(08.09.2 1WGA J1958.2+3232;08.02.1;08.14.2;08.23.01;13.25.5)} 

\title{The nature of 1WGA J1958.2+3232: A new intermediate polar}

\author{I.~Negueruela\inst{1}
\and P.~Reig\inst{2,3} 
\and J.~S.~Clark\inst{4}}                   
                                                            
\institute{SAX SDC, ASI, c/o Nuova Telespazio, via Corcolle 19, I00131
Rome, Italy
\and Foundation for Research and Technology-Hellas, 711 10, Heraklion,
Crete, Greece
\and Physics Department, University of Crete, 710 03, Heraklion, Crete, Greece
\and Astronomy Centre, CPES, University of Sussex, Falmer, Brighton, 
BN1 9QH, U.K.}

\mail{ignacio@tocai.sdc.asi.it}

\date{Received    / Accepted     }

\titlerunning{The intermediate polar 1WGA J1958.2+3232}
\authorrunning{Negueruela et al.}
\maketitle 

\begin{abstract}
We present low and intermediate resolution spectroscopy of the optical
counterpart to the recently discovered pulsating X-ray source 
1WGA J1958.2+3232. The presence of strong \ion{H}{i}, \ion{He}{i} and
\ion{He}{ii} emission lines together with the absence of absorption 
features rules out the possibility that the object is a massive star,
as had recently been suggested. 
The observed X-ray and optical characteristics are consistent 
with the object being an intermediate polar.
The double-peaked structure of the emission lines indicates that
an accretion disc is present.
 \end{abstract}


\section{Introduction}

There are several types of X-ray sources which display significant 
modulation in their X-ray lightcurves, among which isolated neutron
stars, anomalous X-ray pulsars and two types of well characterised
binary systems: accreting X-ray pulsars (accreting neutron stars with
strong magnetic fields $B\ga10^{11}\:{\rm G}$) and magnetic cataclysmic 
variables (accreting white dwarfs with moderate magnetic fields 
$B\ga10^{5}\:{\rm G}$). Recent systematic analysis 
of {\em ROSAT} observations has resulted in the detection of several
new such sources. However, given the limited spectral information
of the {\em ROSAT} data and the impossibility of determining the intrinsic 
luminosity of the sources, the classification of these objects depends 
on the identification of their optical counterpart. X-ray pulsators
are generally part of a high mass X-ray binary (HMXRB) and their optical
spectra are those of the massive companion, without any significant
contribution from the vicinity of the neutron star. In cataclysmic
variables (CVs), on the other hand, the white dwarf is accreting from a 
late-type unevolved star and the optical spectrum is dominated by emission
from the accretion disc (if present) or the accretion stream (when
the magnetic field is too strong to allow the formation of an
accretion disc). In polars (AM Her stars), the magnetic field 
($B\ga5\times10^{6}\:{\rm G}$) is dominant: there is no accretion disc 
and the orbit and spin periods are synchronised. In magnetic CVs with 
weaker magnetic fields (intermediate polars) rotation is not synchronous
and an accretion disc, an accretion stream or both can be present.

Strong modulation (at an 80\% level) was discovered in the X-ray signal
from the {\em ROSAT}\, PSPC source 1WGA~J1958.2+3232 by Israel et al. (1998).
The pulse period was poorly determined at 721$\pm$14 s, though
a later {\em ASCA} observation allowed the derivation of a much more 
accurate value 734$\pm$1 s (Israel et al. 1999). The energy spectrum
was fitted by a simple power law giving a photon index $\Gamma = 
0.8^{+1.2}_{-0.6}$ and a column density $N_{{\mathrm H}} = \left(
6^{+24}_{-5}\right)\times10^{20}\:{\rm cm}^{-2}$. Given these parameters
and the fact that the source was close to the Galactic plane, Israel
et al. (1998) were unable to decide whether the source was a low-luminosity
persistent Be/X-ray binary (see Negueruela 1998; Reig \& Roche 1999) or 
an intermediate polar (see Patterson 1994).

\begin{figure*}[t]
\begin{picture}(500,230)
\put(0,0){\includegraphics{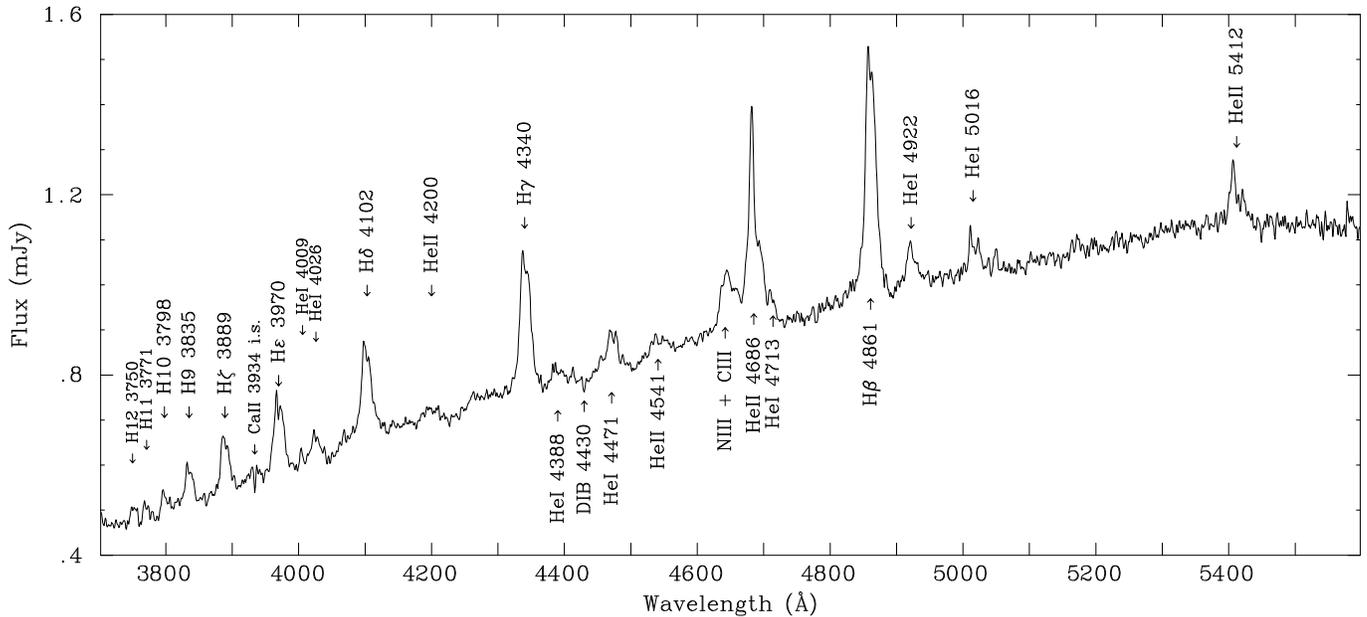}}
\end{picture}
\caption{Blue spectrum of the optical counterpart to  1WGA J1958.2+3232,
taken on July 12, 1999, with the WHT and ISIS. Emission lines are marked.}
\label{fig:blue} 
\end{figure*}

Later Israel et al. (1999) located an $V=15.7$ emission line object 
inside the 30$\arcsec$ X-ray error circle, which is the optical 
counterpart. Based on a low signal-to-noise spectrum, Israel et al.
(1999) classified the object as a Be star, in spite of the evident
presence of strong \ion{He}{ii}\,$\lambda$4686\AA\ emission, which is 
never seen in classical Be stars (in Be/X-ray binaries, if at all present,
it only shows as some in-filling in the photospheric line). 
Based on some features that they
identified as interstellar lines, they speculated that the optical
counterpart was a slightly reddened B0Ve star at a distance of 800 pc.
However, if a B0V star was slightly reddened, it should have an apparent 
magnitude $V \approx 6$ rather than $\approx 16$, and a very large reddening
is unlikely given that the extinction in that
direction has been measured to be small (Neckel \& Klare 1980) . 
This led us to obtain higher resolution
spectra of the source. In this paper we show that the optical properties
of the object clearly identify it as an intermediate polar, rather than
a Be/X-ray binary.

\section{Observations}

\subsection{Optical spectroscopy}

\begin{figure}[ht]
\begin{picture}(250,220)
\put(0,0){\includegraphics{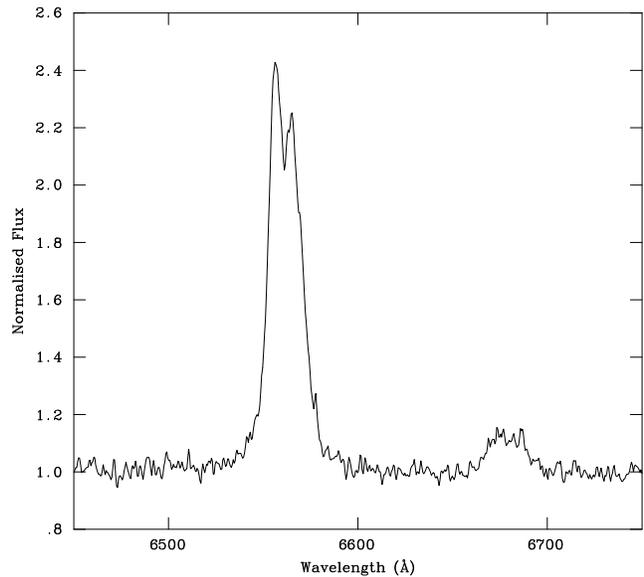}}
\end{picture}
\caption{H$\alpha$ and \ion{He}{i} $\lambda$6678\AA\ emission lines in the
optical counterpart to 1WGA J1958.2+3232. Spectrum taken with the WHT and 
ISIS on July 12, 1999.} 
\label{fig:red} 
\end{figure}

 We observed the optical counterpart to 1WGA J1958.2+3232 on July 12, 1999,
using the Intermediate Dispersion Spectroscopic and Imaging System (ISIS) 
on the 4.2-m William Herschel Telescope (WHT), located at the Observatorio 
del Roque de los Muchachos, La Palma, Spain. The 
blue arm was equipped with the R300B grating and the EEV\#10 CCD, which
gives a nominal dispersion of $\sim 0.9$ \AA/pixel over $\sim 3500$ \AA.
 The red arm was equipped with the R1200R grating and the Tek4 CCD,
which gives a nominal dispersion of $\sim 0.4$ \AA/pixel at H$\alpha$.
The exposure time was $1500\:{\rm s}$. The data were processed using the 
{\em Starlink} packages {\sc ccdpack} (Draper 1998) 
and {\sc figaro} (Shortridge et al. 1997) 
The extracted spectra are displayed in Figures~\ref{fig:blue} and 
\ref{fig:red}.

\begin{figure}[ht]
\begin{picture}(250,120)
\put(0,0){\includegraphics{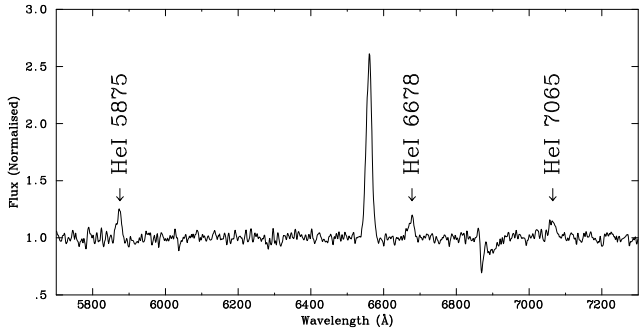}}
\end{picture}
\caption{Low-resolution spectrum of the optical counterpart to 
1WGA J1958.2+3232, taken on July 26, 1999, with the 1.3-m telescope
at Skinakas.} 
\label{fig:lowres} 
\end{figure}

We obtained lower resolution spectroscopy using the 1.3-m Telescope at the
Skinakas Observatory (Crete, Greece) on July 26, 1999. The telescope is an
f/7.7 Ritchey-Cretien and was equipped with a 2000 $\times$ 800 ISA SITe
chip CCD. This camera has 15$\mu$m pixels and reaches maximum efficiency
($\sim$ 90\%) in the red, at around H$\alpha$. The spectrum (a 1800-s
exposure) was taken with
a 1300 line mm$^{-1}$ grating and a 320 $\mu$m width slit (6$''$.7) which
gave a dispersion of 1 \AA \ pixel$^{-1}$. The spectrum, which is 
displayed in Figure~\ref{fig:lowres}, was reduced using
{\sc figaro}.

\subsection{Optical photometry}

We obtained Str\"{o}mgren photometry of the field using the 1.3-m
Telescope at Skinakas Observatory on August 16, 1999 (JD 2,451,407). The
telescope was equipped with  a 1024 $\times$ 1024 pixel  SITe CH360 CCD.
The size of the pixels was 24$\mu$m, representing approximately 0$''$.5 on
the sky. The source was observed through standard $u$, $v$, $b$, $y$
filters with exposure times of 1200, 900, 600 and 300 seconds,
respectively. A sufficient number of standards were observed in order to
compute the atmospheric extinction coefficients and allow the
transformation to the standard system.

The results  are displayed in Table~\ref{tab:phot}. We have also obtained
measurements for the only other star of similar brightness which was inside
both the {\em ASCA} and {\em ROSAT} error circles -- dubbed  ``candidate A'' 
by Israel et al. (1999). As can be seen, the
values of $y$ for the proposed optical counterpart and candidate A 
are compatible with the $V$ values obtained by Israel et al. (1999) -- 
15.7$\pm$0.2 and 15.4$\pm$0.2 respectively.

\begin{table*}[ht]
\caption{Observational details of the optical photometry.}
\begin{center}
\begin{tabular}{lcccc}
\hline
& $y$  &  $b$ & $v$ & $u$\\
\hline
&&&&\\
1WGA J1958.2+3232&15.77$\pm$0.04&15.94$\pm$0.04&16.12$\pm$0.05&16.48$\pm$0.07\\
Candidate A &15.55$\pm$0.05&16.15$\pm$0.05&16.56$\pm$0.07&17.60$\pm$0.16\\
\hline
\hline
\end{tabular}
\end{center}
 \label{tab:phot}
\end{table*}

\section{Discussion}

The blue spectrum of the optical counterpart to 1WGA J1958.2+3232 is 
displayed in Figure~\ref{fig:blue}. The spectrum is typical of a cataclysmic
variable with no obvious absorption stellar features and strong emission
in all Balmer lines (down to H12). The absence of photospheric features
rules out the possibility that 1WGA J1958.2+3232 is a Be/X-ray binary 
-- see, for example, Steele et al. (1999), where it is shown that even for
the Be stars with the strongest emission veiling, the photospheric features
allow spectral classification to the spectral subtype.

In the spectrum of 1WGA J1958.2+3232, on the other hand, as is typical 
in intermediate polars, \ion{He}{ii}~$\lambda$4686\AA\ and the Bowen 
complex are strongly in 
emission. Many other \ion{He}{i} and \ion{He}{ii} transitions are also
in emission. The Balmer lines are all double-peaked and asymmetric with
a stronger blue peak (note that the profile of H$\epsilon$ is modified
by the interstellar \ion{Ca}{ii}~$\lambda$3968\AA\  line). The asymmetry
is still stronger in the \ion{He}{ii} lines and can be seen in the weaker
\ion{He}{i} lines. The centroids of emission lines (determined by 
fitting a single Gaussian to the profile) show no displacement from
the rest wavelength within the resolution achieved. The blue peaks
of the \ion{H}{i} and \ion{He}{ii} lines are displaced by $\sim 
250\:{\rm km}\,{\rm s}^{-1}$.

Figure~\ref{fig:red} displays H$\alpha$ and 
 \ion{He}{i} $\lambda$6678\AA\ at higher resolution. The double-peaked
shape can be seen in greater detail in the H$\alpha$ line. This 
is evidence for the presence
of an accretion disc surrounding the white dwarf. The exact shape of the
lines must depend on the orbital phase at which the observation was taken.
 Given that the X-ray flux is strongly pulsed and
an accretion disc is present, the object must be an intermediate polar.
Therefore the observed X-ray variation should represent the spin period of
the cataclysmic variable or the beat period between the spin and orbital
periods, since it should be an asynchronous system.
The sharpness of the peaks indicates that the 25-min exposure does not 
represent a significative portion of the orbit (otherwise the peaks
would be blurred). 
This is consistent with expected orbital periods of a few hours. 

In the lower resolution spectrum taken two weeks later 
(Fig.~\ref{fig:lowres}), H$\alpha$ and the \ion{He}{i} are single-peaked
and red-dominated, indicating that the source was observed at a different 
orbital phase. Even though the resolution is rather lower than in the
WHT spectrum, a peak separation similar to that measured in the first 
spectrum ($v\approx375\:{\rm km}\,{\rm s}^{-1}$) should have been 
resolved. The interstellar \ion{Na}{i} lines are not detectable above 
the noise
level. Due to their weakness and the irregularity of the continuum, no 
diffuse interstellar bands (DIB) can be measured even in the higher
resolution spectra. We set upper limits for the Equivalent Width (EW) of 
the DIBs at $\lambda$4430\AA\ and $\lambda$6613\AA\ as 
EW$<400\:{\mathrm m}$\AA\ and $<50\:{\mathrm m}$\AA, both of which are
consistent with $E(B-V)<0.2$  (Herbig 1975). This is in accordance with the
measurements of interstellar absorption in this direction ($l=69\deg,
b=1.7$) by Neckel \& Klare (1980), who find $A_{V} < 0.5\:{\rm
mag}$ and  $A_{V} < 1.0\:{\rm mag}$ at 1 kpc for the two fields between
which  1WGA J1958.2+3232 approximately lies.
 
\begin{figure}[ht]
\begin{picture}(250,120)
\put(0,0){\includegraphics{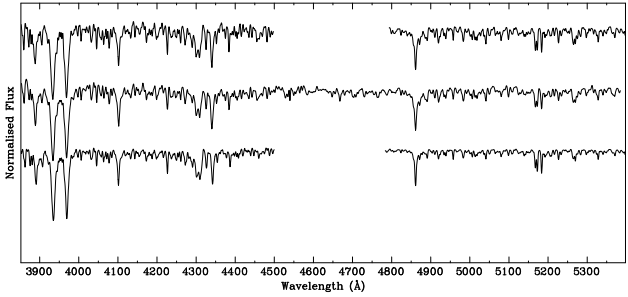}}
\end{picture}
\caption{Spectrum of the star called Candidate A in Israel et al. (1999),
which is only $\sim 40\arcsec$ from 1WGA J1958.2+3232. The comparison spectra
correspond to HD 5015 (top, F8V) and HD 22879 (bottom, F9V) and are taken 
from the database of Leitherer et al. (1996).} 
\label{fig:canda} 
\end{figure}

In the WHT observations, we set the slit in
such a way as to also observe the nearby star dubbed ``Candidate A'' by 
Israel et al. (1999), which is about $40\arcsec$ away from the optical
counterpart to 1WGA J1958.2+3232, and therefore could provide some 
information on the reddening in that direction. Even though 
Israel et al. (1999) claim that this object is
an early-type star, comparison with the spectra of several stars taken
from the electronic database of Leitherer et al. (1996) shows that its
spectral type is F8V (see Fig.~\ref{fig:canda}). We cannot see the 
$\lambda$4430\AA\ DIB down to the level of the many weak features in the
spectrum, which gives an upper limit of ${\rm EW} \approx 300\:{\mathrm
m}$\AA. From the measured  $(b - y) = 0.59\pm0.07$ and the intrinsic
$(b-y)_0=0.350$ for an F8V star (Popper 1980) we obtain the interstellar
reddening $E(b-y)=0.24$. Using the relation of Crawford \& Mandwewala 
(1976) $E(B-V)=1.35 E(b-y)$, this implies $E(B-V)=0.32$, significantly
 higher than the upper limit that could be derived from the interstellar
$\lambda$4430\AA\ DIB, which implies $E(B-V)< 0.13$, according to the 
relation by Herbig (1975). Assuming $M_{V} = +4.2$ for a main-sequence 
F8 star (Deutschman et al. 1976) and the standard reddening $R=3.1$, this 
star is situated at a distance $d \approx 0.9\:{\rm kpc}$.

Given its brightness, 1WGA J1958.2+3232 should be located at a distance
$1\la d \la 1.5\:{\rm kpc}$ (see Israel et al. 1998), i.e., farther away
than the F8V star and therefore would have a higher reddening. 
If the reddening is $E(B-V)>0.3$, the soft X-ray flux could be absorbed, 
which would explain the relatively low $L_{{\rm x}}/L_{{\rm opt}}$ of 
the source when compared to less distant intermediate polars (see 
Israel et al. 1998). We note that the interstellar lines indicate
a lower reddening, but in the F8V star this estimate is also rather
lower than the photometric determination of the reddening.
 
With a pulse period of 
$734\:{\rm s}$, this system falls in between the two groups of
short and long period intermediate polars defined by Norton et al. (1999),
and characterised by different X-ray pulse shapes. Clearly further X-ray
observations of the source are needed and either {\em RXTE} or {\em Chandra}
could provide more detailed timing observations. Also, future time-resolved
photometric and spectroscopic observations are needed in order to determine
the orbital period and whether the observed X-ray pulsations correspond to
the spin period.

\section{Conclusions}

Based on intermediate-resolution spectroscopy, we conclude that 
1WGA J1958.2+3232 is an intermediate polar, rather than a Be/X-ray 
binary. From the magnitudes measured for the object and a very nearby
F8V star we can estimate that 1WGA J1958.2+3232 is situated at a 
distance of 1\,--\,1.5 kpc and moderately reddened with $E(B-V) \la 0.3$. 

\section*{Acknowledgements}

The WHT is operated on the island of La Palma by the Royal Greenwich
Observatory in  the Spanish  Observatorio del Roque de Los Muchachos of
the Instituto de Astrof\'{\i}sica de Canarias. The observations were taken
as part of the ING service observing programme. Skinakas Observatory is a
collaborative project of the University of Crete, the Foundation for
Research and Technology-Hellas and the Max Planck Institut f\"ur
Extraterrestrische Physik. The authors would like to thank Dr. GianLuca
Israel for his help with this work and Drs D.~di~Martino  and A.~J.~Norton
for their helpful comments on the draft. We are also grateful 
to Drs E.~V.~Paleologou and
I.~E.~Papadakis for helping with the spectroscopic and photometric
observations at Skinakas Observatory, respectively. IN is supported by an
ESA external fellowship. PR acknowledges partial support via the European
Union Training and Mobility of Researchers Network Grant
ERBFMRX/CT98/0195.  JSC is supported by a PPARC research assistantship.

\end{document}